\ifx\mnmacrosloaded\undefined \input mn\fi

%

\newif\ifAMStwofonts

\ifCUPmtplainloaded \else
  \NewTextAlphabet{textbfit} {cmbxti10} {}
  \NewTextAlphabet{textbfss} {cmssbx10} {}
  \NewMathAlphabet{mathbfit} {cmbxti10} {} 
  \NewMathAlphabet{mathbfss} {cmssbx10} {} 
  \ifAMStwofonts
    \NewSymbolFont{upmath} {eurm10}
    \NewSymbolFont{AMSa} {msam10}
    \NewMathSymbol{\upi}     {0}{upmath}{19}
    \NewMathSymbol{\umu}     {0}{upmath}{16}
    \NewMathSymbol{\upartial}{0}{upmath}{40}
    \NewMathSymbol{\leqslant}{3}{AMSa}{36}
    \NewMathSymbol{\geqslant}{3}{AMSa}{3E}

     \let\ge=\geqslant
  \else
    \def\umu{\mu}
    \def\upi{\pi}
    \def\upartial{\partial}
  \fi
\fi


\pageoffset{-2.5pc}{0pc}

\loadboldmathnames



\pagerange{1--7}    
\pubyear{1989}
\volume{226}
\input psfig
\begintopmatter  

\title{Infrared Spectra of Cooling Flow Galaxies}
\author{W. Jaffe}
\affiliation{Leiden Observatory, PO Box 9513, 2300RA Leiden, The Netherlands}

\author{M.N. Bremer}
\affiliation{Department of Physics, Bristol University, H. H. Wills Laboratory, Tyndall Ave.,
Bristol, BS8 1TL, England}
\author{P.P. van der Werf}
\affiliation{Leiden Observatory}
\shorttitle{Infrared Spectra of Cooling Flow Galaxies}


\acceptedline{Accepted 1988 December 15. Received 1988 December 14;
  in original form 1988 October 11}

\abstract {We have taken K-band spectra covering 
7 cooling flow clusters.  The spectra show many of the 1-0S
transitions of molecular Hydrogen, as well as some of the higher
vibrational transitions, and some lines of ionized Hydrogen.
The line ratios allow us to conclude that the rotational states of
the first excited vibrational state are in approximate LTE, so that
densities above $10^5$ cm$^{-3}$ are likely, but there is evidence
that the higher vibrational states are not in LTE.  The lack
of pressure balance between the molecular gas and the ionized components
emphasizes the need for dynamic models of the gas.  The ratios
of the ionized to molecular lines are relatively constant
but lower than from starburst regions, indicating that 
alternative heating mechanisms are necessary.
}

\keywords {cooling flows -- ISM:molecules -- dust,extinction}


\maketitle  

\section{Introduction}
Cooling flows result when the hot, 
X-ray emitting, gas in a cluster is
compressed by the gravitational force of the central galaxy to a density
where it can cool in less than the Hubble time (c.f. Fabian 1994).
In many cases 100-1000 M$_\odot$/yr condense out of the hot gas.
They present at least two major unsolved problems:
the ultimate fate of the gas after 
it has cooled, and the source of the copious optical line emission
in the centre of the flows (Heckman et al. 1989, 
Voit and Donahue, 1997: VD97).

A new aspect of both these problems was revealed by the detection of
infrared line emission of H$_2$ molecules from the centres of several
CF clusters (Mouri 1994,
Elston and Maloney 1994, Jaffe and Bremer 1997 (Paper I),
Falcke et al. 1998, Genzel et al. 1998).  
This emission arises from dense gas at $\sim
2000$ K and is not found in similarly radio-loud AGNs not associated
with CFs.  Thus it represents {\it prima facie} evidence for ``cool"
material in cooling flows.  Its spatial extent and energetics suggest
an intimate association with the optical line emitting gas (Paper
I). High surface brightness molecular line emission is generally
concentrated within the giant central cluster galaxies (typically in
the inner 10 kpc) in a similar manner to the emission from ionised
gas.  Both can be considered galactic-scale phenomena, even though the
presence of the gas is statistically correlated with the presence of a
cooling flow on larger scales (Paper I). 

We do not yet understand the heating mechanism of the molecular gas,
and cannot accurately estimate the mass associated with it.
The mass in gas {\it at} 2000 K is only about 10$^{5-6}$ M$_\odot$ but
the cooling time from this temperature is very short, about one year.
Without a consistent thermal model, we do not know whether this
represents a minor mass component (the equivalent of only 1000 years
of CF mass accumulation) that is kept continually warm, or the tail of
a cooler but much more massive molecular dog that is heated to IR
temperatures by shocks, X-rays, or stellar photoionization.

In this paper we describe new K-band IR spectroscopy of six cooling flows
taken with the United Kingdom InfraRed Telescope (UKIRT) on 2 and
3 January, 1997.
By studying the ratios between the various lines of H$_2$ and between
the H$_2$ and HII lines we hope to shed light on the thermal state
of the gas, and ultimately on the heating mechanism.
The clusters studied were Abell 0335, 478, 1795, and 2029, Hydra A, NGC 1275, 
and PKS~0745-19.  They were observed with the CGS4 detector in
the following configuration: 75 l/mm
grating; 256(spectral)$ \times$ 84(spatial) InSb array; slit size:2",
spatial scale 1"/pixel, dispersion: .0026$\mu$/pixel.
\section{Observation and Reduction}
The observations followed standard IR spectroscopic techniques.  
Each obervation consists of a series of 60s frames, 
consisting of four subframes, each shifted 1/2 pixel 
in the spectral direction.  The four subframes are combined into a single
514x84 frame.  

We calibrated the atmospheric transmission by observing, with
the same technique, standard stars about every 90 minutes.
We chose the stars from the Carlsberg Meridian Catalog 
(Fabricius, 1993) to be of early F type and within 10$^{\rm o}$ of 
the target galaxies.  

For any 1-dimensional target spectrum we used the telluric calibrators to 
subtract the smooth underlying stellar continuum leaving only
non-atmospheric absorption and emission features.  
Our procedure is to find emission lines in these spectra,
and then correct the fluxes for the atmospheric absorption, rather than to
first correct the spectra and then fit lines.  For narrow lines,
the former procedure is less sensitive to spurious lines created
by minor atmospheric fluctuations.  
These steps are illustrated for the cluster PKS0745-11 in Figure 1.
\beginfigure{1}
\psfig{figure=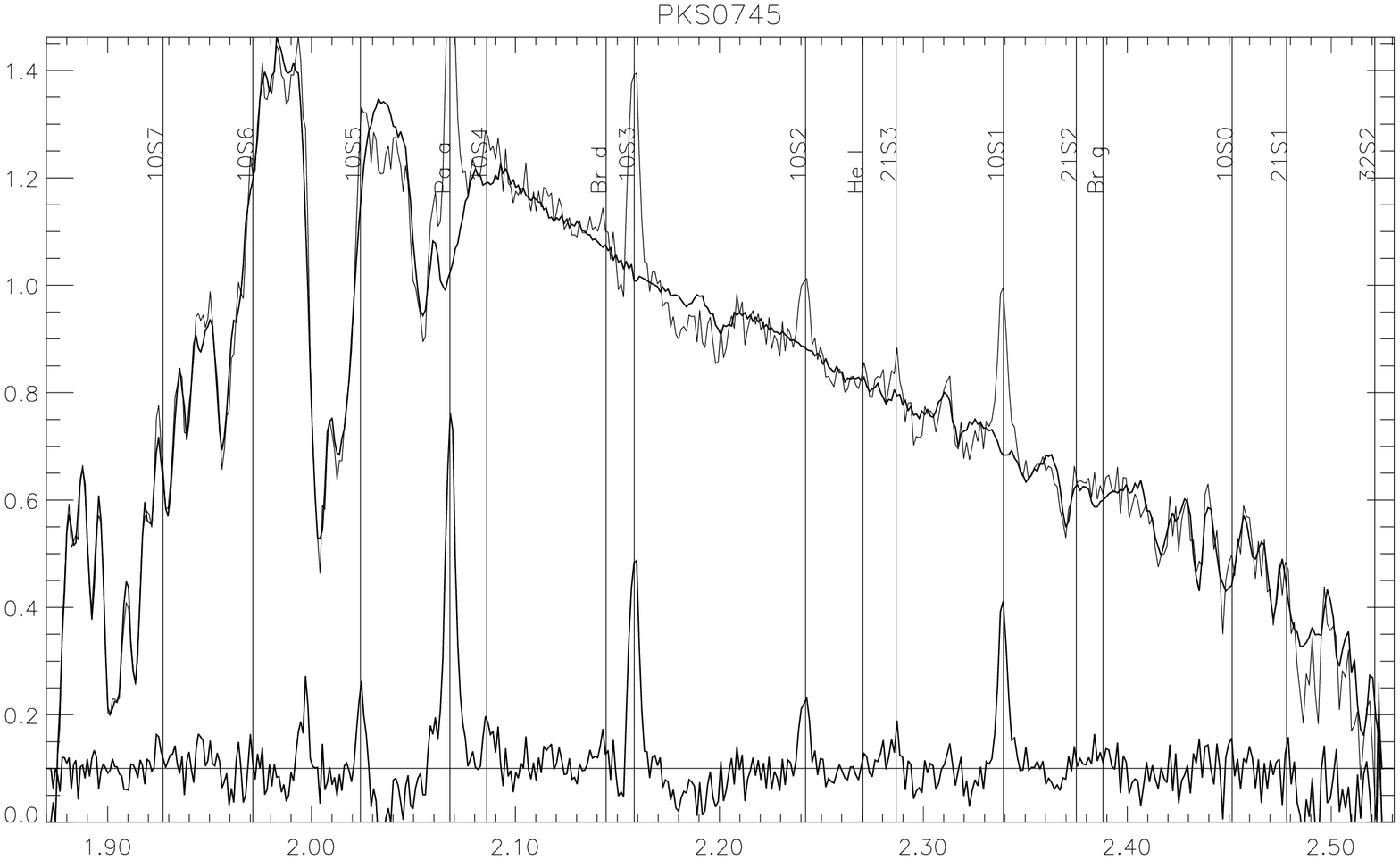,angle=90,height=5truecm}
\caption{{\bf Figure 1} The top section shows the extracted spectrum
and the best fitting continuum fit, including telluric absorption.
The bottom section shows the difference -- the uncorrected emission 
profile.  Wavelengths are in the telescope system}
\endfigure

We applied this procedure to the summed 2-dimensional spectra for
each cluster as follows:  A continuum model, with atmospheric
corrections, was subtracted from each row of the spectrum.
This yields a 2-dimensional spectrum with emission/absorption lines only.
We then summed the spatial profiles of the strongest hydrogen emission
lines. The resulting profile was compared to the spatial profile
of temporally adjacent standard stars to determine the spatial extent of the
emitting regions.  Lastly the rows of the 2-dimensional emission
line spectrum were weighted by this spatial profile and summed to
yield a single profile with the optimum signal/noise for detecting
weak lines, assuming that all lines have the same profiles.

From the uncorrected spectra we determined, by gaussian fitting, the line
widths of the strongest H$_2$ lines.  We then searched for weaker
lines by fitting gaussians of this given width at the given wavelengths
of a large number of H$_2$, HII, and metal lines.  
Because the position and width of the gaussians were fixed, 
the fitting procedure
should have no bias toward finding positive peaks.  We estimated the
uncertainties in the fitted fluxes empirically by applying
the same procedure at 100 random points along the spectrum and
producing a smoothed r.m.s. average of the fluxes so determined
as a function of wavelength.  The method
method includes residual atmospheric features and usually indicates
a larger uncertainty in regions near strong telluric lines.

As a last step, the line strengths are converted to physical units
by correcting for atmospheric absorption and using the standard
star magnitudes.  In this process an estimate of the uncertainty in 
the absorption is added in quadrature to the uncertainty in the
gaussian estimate.  The resulting flux and error estimates are given
in Table 1 for each cluster except Abell~2029 where no
lines were detected during 12 min of observation.  The error estimates 
do not include the uncertainty of the absolute flux scale.  This
last uncertainty is relatively large because few of the Carlsberg
calibrators have known K-band magnitudes, so their magnitudes were
extrapolated to K-band from known B and R fluxes using a standard
F-star spectrum.  We estimate the additional uncertainty in the 
absolute line flux scale to be of order 30\%, but the relative
line strengths within one spectrum, or in one cluster relative to another,
are accurate at the levels quoted in Table 1.
\beginfigure{2}
\psfig{figure=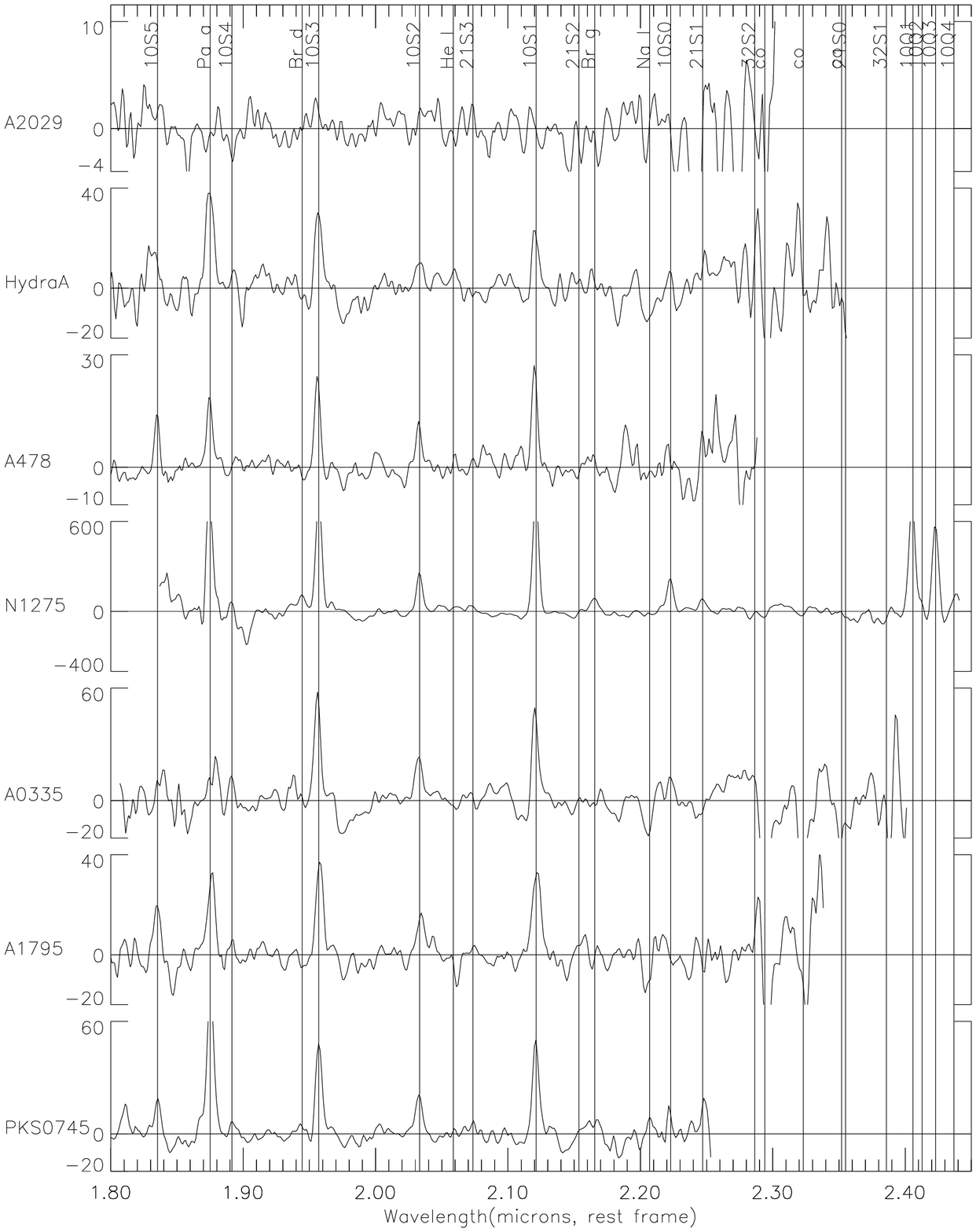,height=12truecm}
\caption{{\bf Figure 2.} Calibrated and continuum subtracted spectra
for the seven clusters studied.  Wavelengths are in $\mu$ in the
rest frame and the vertical scales are in 10$^{-16}$ erg 
cm$^{-2}$ s$^{-1}$ \AA$^{-1}$
}
\endfigure
\begintable*{1}
\caption{{\bf Table 1.} Line fluxes from the studied clusters.  Fluxes
are in units of 10$^{-16}$ erg cm$^{-2}$ s$^{-1}$.  }
\halign{%
#\hfil&\qquad\hfil#&\qquad\hfil#&\qquad\hfil#&\qquad\hfil#&\qquad\hfil#&\qquad\hfil#\cr
Cluster&PKS0745&A0335&NGC1275&Abell 1795&Abell 478&Hydra A\cr
Exposure Time (min)&60&54&12&30&100&56\cr
z&0.1028&0.0349&0.0017&0.0616&0.0860&0.0540\cr
1-0S(0) (2.2235$\mu$)&$3.4\pm5.9$&$6.4\pm1.6$&$102.0\pm6.0$&$-0.6\pm1.4$&$2.4\pm 1.4$&$3.7\pm1.3$\cr
1-0S(1) (2.1218$\mu$)&$24.4\pm0.7$&$22.1\pm1.1$&$384.3\pm3.8$&$15.8\pm1.0$&$13.3\pm 0.8$&$ 9.7\pm0.7$\cr
1-0S(2) (2.0338$\mu$)&$ 9.2\pm0.6$&$ 9.9\pm0.8$&$115.1\pm6.2$&$ 7.4\pm0.6$&$ 5.9\pm 0.6$&$3.1\pm0.6$\cr
1-0S(3) (1.9576$\mu$)&$22.8\pm$0.5&$25.7\pm1.9$&$332.3\pm14.$&$17.4\pm0.4$&$12.0\pm 0.4$&$15.5\pm1.6$\cr
1-0S(4) (1.1.8920$\mu$)&$3.2\pm0.5$&$6.4\pm11.$&$59.2\pm28.$&$3.2\pm8.4$&$-1.4\pm 0.8$&$2.9\pm4.3$\cr
1-0S(5) (1.8358$\mu$)&$9.5\pm0.8$&N.A.&N.A.&$9.1\pm5.2$&$6.9\pm1.3$&$6.2\pm7.7$\cr
2-1S(1) (2.2477$\mu$)&$7.2\pm6.0$&$-6.3\pm3.0$&$34.7\pm6.$&$-0.9\pm1.0$&$3.3\pm 4.3$&$3.2\pm2.1$\cr
2-1S(2) (2.1542$\mu$)&$1.4\pm0.8$&$-0.1\pm1.0$&$10.9\pm4.9$&$3.8\pm1.2$&$-1.6\pm 0.7$&$1.3\pm0.9$\cr
2-1S(3) (2.0735$\mu$)&$3.1\pm0.6$&$-0.4\pm0.8$&$13.1\pm3.6$&$2.6\pm0.9$&$0.5\pm 0.7$&$-2.8\pm1.0$\cr
Pa $\alpha$ (1.8756$\mu$)&$36.3\pm0.7$&$5.0\pm12.$&$300^1$&$14.0\pm3.1$&$8.4\pm 0.6$&$17.6\pm1.3$\cr
Br $\gamma$ (2.1661$\mu$)&$3.7\pm0.8$&$-0.6\pm1.0$&$47.8\pm4.7$&$0.4\pm1.3$&$2.0\pm 0.9$&$1.3\pm1.0$\cr
Br $\delta$ (1.9451$\mu$)&$1.7\pm0.5$&$-2.\pm12.$&$40.1\pm7.7$&$-1.7\pm1.0$&$-0.8\pm 0.4$&$-1.8\pm1.9$\cr
}
\tabletext{$^1$ The Pa$\alpha$ line in NGC~1275 is clearly visible in
Fig. 2, but lies in a region of strong and variable atmospheric absorption.
The true line flux is highly uncertain.}
\endtable
\section{Reduced Spectra and Derived Quantites}
In Figure~2 we present the continuum subtracted and absorption
corrected spectra for all clusters in physical units and in the rest
wavelength frame.  
In Table 2 we present for each cluster the measured velocity dispersion
of the brightest H$_2$ lines, the radial extent of the emission (standard deviation of
gaussian fit along slit, deconvolved with stellar profile), 
the total luminosity in the detected
H$_2$ lines the luminosity in Pa$\alpha$, luminosity in H$\alpha$ (Heckman et al., 1989; White et al. 1994 for
Abell 478) and the excitation temperature $T_{5:3}$,determined from the
ratio of the 1-0S(3) and 1-0S(1) lines (c.f. Section 4.2).  The instrumental
dispersion, measured on arc and sky lines was $240\pm 10$ km s$^{-1}$.  This
may be somewhat larger than the effective dispersion for emission that
does not uniformly fill the slit.  In comparing our luminosities with
those of Heckman et al., determined by narrow band imaging, our 
luminosities should be corrected for light blocked by the slit jaws. This
can be crudely estimated by assuming circular symmetry and calculating
the light lost in this case from the emission profile along the slit.  
This leads to corrections ranging from
1.6 to 2, but these will be incorrect if the source is highly asymmetric,
or if we have missed a very large, low brightness component.  The
latter possibility is particularly likely in the nearby cluster NGC~1275.
\begintable*{2}
\caption{{\bf Table 2. Quantities derived from analysis of the spectra.}
}
\halign{
#\hfil&\qquad\hfil#&\qquad\hfil#&\qquad\hfil#&\qquad\hfil#&\qquad\hfil#&\qquad\hfil#&\qquad\hfil#\cr
Cluster&r(kpc)&$\sigma_v$ (km s$^{-1}$)&$L_{H_2}$ (erg s$^{-1})$&$L_{{\rm Pa}\alpha}$&$L_{{\rm H}\alpha}$&$T_{5:3}$\cr
PKS0745&1.5&320&$1.7\ 10^{41}$&$7.3\ 10^{40}$&$1.3\ 10^{42}$&$1771\pm 200$\cr
A0335&0.5$^1$&370&$1.5\ 10^{40}$&N.A.&$5\ 10^{41}$&$2446\pm 900$\cr
NGC1275&0.13$^2$&300&$6.1\ 10^{38}$&N.A.&$2.1\ 10^{41}$&$1628\pm 100$\cr
Abell 1795&1.1&280&$4.2\ 10^{40}$&$1.0\ 10^{40}$&$4.4\ 10^{41}$&$2217\pm 350$\cr
Abell 478&1.8&260&$5.8\ 10^{40}$&$1.2\ 10^{40}$&$1.7\ 10^{41}$&$1693\pm 362$\cr
Hydra A&1.2&370&$2.3\ 10^{40}$&$9.5\ 10^{39}$&$7.2\ 10^{40}$&$^31480 \pm 300$\cr
}
\tabletext{
$^1$A0335--gaussian fit to spatial fit is poor.  There is an unresolved core with $80\pm10$\% of the
flux.  The remaining 20\% extends to at least 3 kpc and has a gaussian fit radius of about 2 kpc.
$^2$NGC~1275--gaussian fit is poor.  Unresolved core with $\sim 85$\% of flux.  Remainder extends
$\sim$ 10" or 3 kpc in radius.
$^3$ see text at end of section 4.2
}
\endtable
The velocity dispersions, corrected for instrumental widening, run from
100 to 300 km s$^{-1}$.  These are comparable to the dispersions measured
by Heckman et al. (1989) in H$\alpha$ and are much smaller than the
typical velocities in the clusters, the stars in the envelopes
of cD galaxies or gas in the narrow or broad line regions around
active nuclei.

\section{Discussion}
\subsection{Optical Extinction}
The ratio of H$\alpha$ to Pa$\alpha$ fluxes should measure the
extinction of the optical light.  This ratio, for Case B HII regions,
should be about 7 (Osterbrock, 1989), and values much below this would
indicate dust around the emission regions.  Unfortunately the fluxes
are measured with different techniques: slit spectrocopy for
Pa$\alpha$ and narrow band imaging for H$\alpha$.  Hence the ratio is
subject to uncertainties in the absolute calibration, and the flux
lost outside of the spectroscope slit.  For the four high
redshift clusters with good Pa$\alpha$ fluxes, we have taken the ratio
$L_{H_\alpha}/2 L_{Pa\alpha}$, where the factor of 2 corrects for slit
losses.  The values of the ratio are 10, 22, 7, and 4, with an average
of 11.  While the variation confirms the difficulties with
calibration, we conclude that there is no evidence for strong
extinction of the H$\alpha$ fluxes (e.g. more than 1 magnitude).

\subsection{LTE modelling of H$_2$ line ratios}
If the warm gas is sufficiently dense that collisional
transitions dominate over radiative transitions from the
relevant states, then we can regard the gas to be in local
thermal equilibrium (LTE).  We regard this case as a reference model
and ask if the measured line ratios are consistent with this model.
In Table 2, we have given the excitation temperatures derived from
the ratio of 1-0S(1) and 1-0S(3) lines;
in Figures 3 we plot for each cluster $\log(F/ag)$ versus
$T_u$.  Here $F$ is the measured flux in each line, 
and $a$, $g$ and $T_u$ are 
the Einstein coefficient, statistical weight and energy (expressed
as a temperture, $T_u=E_u/k$) of the upper state of the transition.
In LTE, of course, $\log(F/ag)=Constant\ -T_u/T$. 
\beginfigure{3}
\psfig{figure=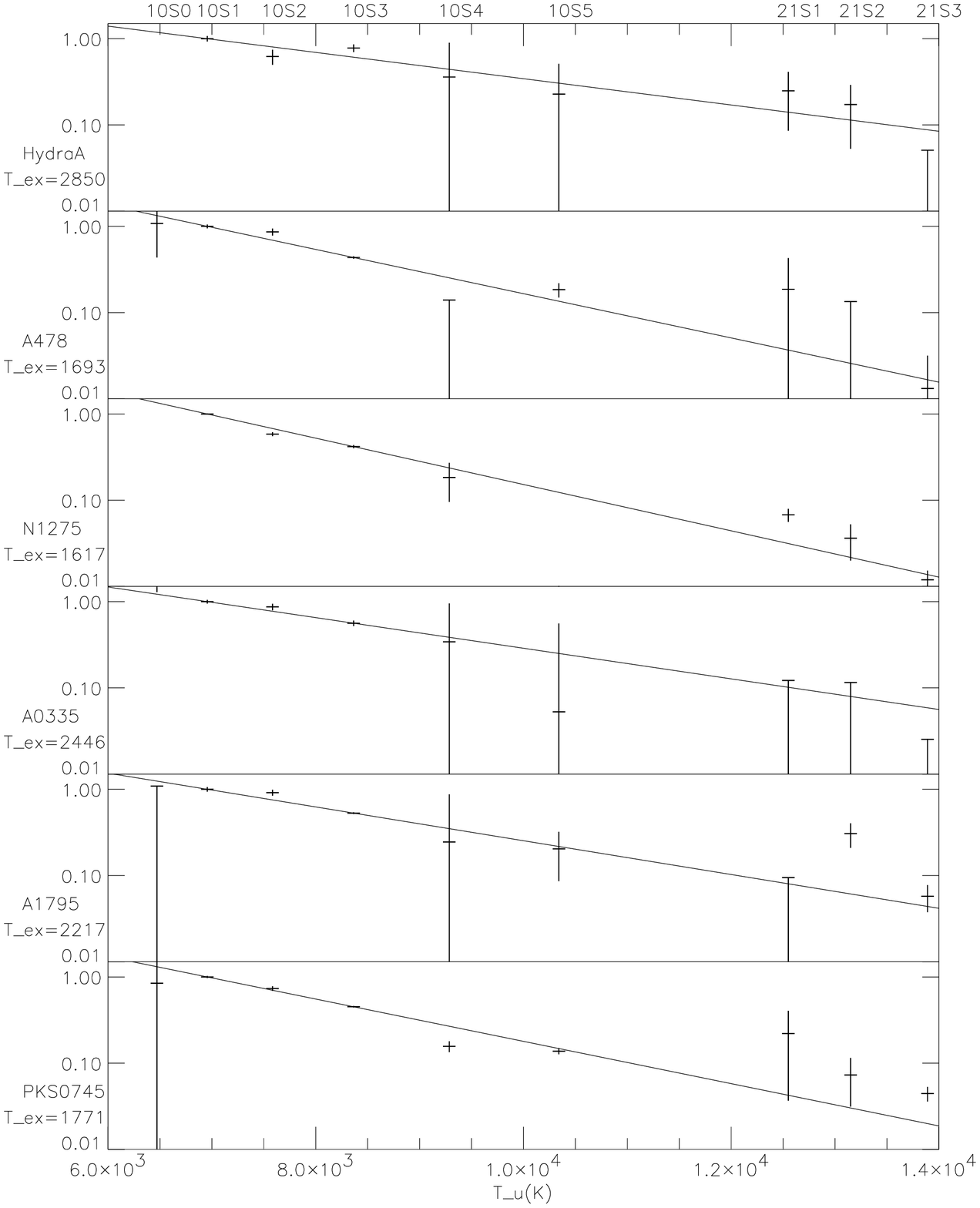,height=12cm}
\caption{Figure 3.  LTE Diagrams for the six clusters with detected lines.
For each cluster the plot shows $\log F/ag$ vs. $T_u$ with the
straight line through the 1-0S(1) and 1-0S(3) points.  
For undetected lines, 2$\sigma$ upper limits are given.
}
\endfigure

From Table 2 and Figure 3 we conclude that the LTE model is
a reasonable approximation for the 1-0 lines in all clusters, 
with excitation temperatures of $\sim 2000$ K.  
There are some lines which deviate significantly
more than the formal errors and these we have investigated individually.
In some cases we have downgraded the significance of the result, because
the line appears in a spectral region with a poor baseline.  This
is unfortunately true for many of the high excitation, 
but quite weak, 2-1 lines.
By this standard, the remaining significant deviations from LTE are
as follows:
{\parskip = \baselineskip\parindent=0pt
\item{} For NGC~1275, the 2-1S(1) lies a factor of 3 above the LTE model
for 1617 K.  This is the cluster with the highest signal/noise ratios.
Krabbe et al. (2000) discuss deeper observations of this object,
where 2-1S(1),S(3) and 3-2S(3) are clearly detected.  They find
the higher lines to require higher excitation temperatures: 
2600 K versus 1500K for the 1-0 lines) and
attribute this largely to fluorescent excitation by Ly$\alpha$ photons.
They conclude that for most transitions collisional processes
dominate, requiring $n_e>10^5$ cm$^{-3}$.
\item{} For A478 and PKS0745, and marginally for N1275 the 1-0S(4)
line lies about a factor of 2 below the model.
\item{}  For Hydra~A the ratio of 1-0S(3) to 1-0S(1) is so high (1.9) that no
plausible excitation temperature can be assigned.  This is probably
due to the fact that the 1-0S(3) line falls in a telluric absorption
band at 2.06$\mu$ and the correction for this is inaccurate.  Excluding
this line, the estimated excitation temperature for all the other
1-0S lines is $1480\pm 50$ K
\item{}
}

\subsection{Physical Implications of the H$_2$ emission}

We conclude from the above analysis that for the clusters studied the
rotational states of the first excited vibrational state are close to
LTE.  This implies a gas density high enough that collisional
transitions state dominate over radiative transitions.  
Using collision cross sections from Allison \& Dalgarno
(1967) and Sternberg \& Dalgarno (1989), modified using more recent
results by Mandy \& Martin (1993), at $T=2000\;$K a molecular hydrogen
density $n\ge 10^6\;$cm$^{-3}$ is required for a 84\% thermal
population of the $v=1$, $J=7$ level which decays through the H$_2$
$v=1-0$ S(5) line detected in A478 and PKS0745.
For the remaining clusters, the highest line
detected is H$_2$ $1-0$ S(3) which is 83\% thermalized at
$n=3\cdot10^5\;$cm$^{-3}$. Using collision cross sections from Le
Bourlot et al (1999) would even increase these densities somewhat. On
the other hand, if there is a significant electron abundance in the
molecular gas, collisions with electrons would contribute to the
excitation, decreasing the required densities. With a fractional
ionization of 1\%, the required densities would be decreased by a
factor of two, for a fractional ionization of 10\% the densities would
be decreased by a factor of 10. Based on these considerations it is
safe to conclude that the densities of the emitting regions are at
least $10^5\;$cm$^{-3}$, and possibly even higher.  At these
densities, the line ratios provide little information on the gas
heating source, since the signature of non-thermal excitation
processes such as UV-pumping or X-ray excitation is quenched, except
in lines of very high critical density which are faint and not
detected in the present data.

The high density and temperature of the gas emitting the near-IR lines
imply high thermal pressures ($n T\sim 10^{8-9}$ cm$^{-3}$K),
exceeding by of the order 100 to 1000 that of the X-ray gas in
the region near the galaxy ($T\sim
10^7$, $n_e\sim 10^{-1}$) or the ionized optical line emitting gas
($T\sim 10^4$, $n_e\sim 10^2$) as estimated from the [SII] doublet
ratio (Heckman et al., 1989).  The latter two phases have
comparable pressures.  Because of this,
most models of the {\it optical} line emission in
the literature assume pressure equilibrium with the X-ray phase. 
Some theoretical work has been done on the expected properties
of the molecular phase in cooling flows (e.g. Fabian 1994).
Our results make it evident that a dynamic, non-isobaric model must be
used to self-consistently explain the ionized and molecular line emission.

One possible model suggested by the obvious pressure imbalance, and the
strong linkage of the optical and K-band fluxes, is an ablative or
"rocket" model of the gas system.  X- or UV-radiation could heat the
surface layers of a cold, gravitaionally bound, molecular cloud to
2000 K.  At this temperture the gas is no longer gravitationally bound
and will expand into the surrounding space.  As the density decreases
the material will ionize and emit the observed optical line emission.
We will explore this model, and test the predicted line ratios, in a
future paper.

\subsection{Relation of HII to H$_2$ emission}

In general, the H$_2$ and HII emission in cooling flows 
are similarly distributed. As noted in Paper I, the H$_2$ and HII
emission lines correlate well spatially, energetically, and
kinematically.  An extreme example of this is Abell 2029, which
to date has shown no detectable H$\alpha$ or H$_2$ IR lines, although
it is a strong cooling flow and strong radio source.  
In the two nearest clusters, NGC1275 and A0335, 
the IR line emission is dominated by
an unresolved, presumably nuclear component. This is partly
an artifact of their large angular size which excludes some
of the extended emission from the slit.  Both of these cluster
have in fact, an extended component, as do the spectra of all of the other
objects. 

More recently, two of the objects in 
our sample were imaged with H$_2$ lines at high resolution with
HST/NICMOS by Donahue et al., (2000). In the
images, the H$_2$ emission traces H$\alpha$
emission. This implies that the H$_2$ and HII gas are cospatial on
scales of a few hundred parsecs or less and strongly suggests that the
excitation mechanisms of the two gas phases are either the same, or
closely coupled. 

For the four clusters in Table 2 for which good Pa$\alpha$ fluxes are 
available, the ratios of the ratios of $L_{H_2}/L_{Pa\alpha}$ 
range from 2.3 to 4.8 (mean=3.4).  The single line ratio $L_{Pa\alpha}/
L_{1-0S(1)}$ ranges from 0.6 to 1.8 (mean 1.0; rms 0.4).
Br$\gamma$ is generally not detected; the estimated ratio to 1-0S(1) is
less than 0.15 in all cases.

The ratios of the HII lines to the H$_2$ lines are lower than those
seen in ``active" comparison galaxies: Seyferts, starburst galaxies
and Ultraluminous InfraRed galaxies (ULIRGs).  
For all these galaxies, the ratio
of Br$\gamma$/1-0S(1) is greater or equal than 1 (
Moorwood \& Oliva, 1988, 1990, Puxley et al., 1988,
Genzel et al. 1995, Goldader et al., 1995,
Thompson 1995, Vanzi et al., 1998,
Genzel et al., 1998, Schinnerer et al., 1998,
Murphy et al., 1999).  We discuss the implications of this difference for 
UV-photoionization and shock excitation models below.

From the consistency of the line ratios, the spatial overlap, 
and the difference of the ratios from those of other active
galaxies, we conclude that in most cooling flows the 
excitation of radiation from the ionized and
molecular gas are causally related.  This may not be true for
some galaxies with exceedingly active nuclei.  
Cygnus-A for example shows a much higher ratio of Pa$\alpha$/1-0S(1) 
than we find (Wilman et al., 2000).

\subsection{Excitation by stellar UV radiation}
Some authors (e.g. VD97) have suggested that the
optical spectra of cooling flows can be explained largely by stellar
photoionization.  This cannot be the case for the molecular gas
detected here.  The faintness of Br$\gamma$ in the
cooling flows, compared to starburst galaxies,
argues against H$_2$ excitation by star formation.
Photo-ionization by the hottest stars is also known to
be insufficient to account for the high electron temperatures in the
ionized filaments of cooling flows, which thus already required an
additional source of excitation (VD97). The addition
of the IR spectra to the optical spectra then potentially requires
{\it three} excitation mechanisms to be acting simultaneously, and in
concert, to yield the uniform IR spectra presented
here. 

An additional diagnostic is provided by the ortho/para ratio, which is
expected to be three (the ratio of the statistical weights) in situations
where the high-temperature equilibrium ortho/para ratio is reached
(Burton et al. 1992). On the other hand, in regions where the H$_2$
lines are UV-excited, a significantly lower ortho/para ratio is
expected, due to spin-exchange reactions with protons and hydrogen
atoms (Draine \& Bertoldi 1996). Indeed, a low ortho-para ratio is
observed in the extended H$_2$ vibrational emission in the starburst
galaxy NGC\,253 (Harrison et al.\ 1998) and in regions of the
well-studied Galactic photodissociation region in the Orion bar
(Marconi et al., 1998) and the giant extragalactic HII
region NGC~5461 (Puxley et al., 2000). The low
ortho-para ratio in UV-excited regions has been highlighted
in the models of Sternberg \& Neufeld (1999).

In all of our targets, the $v=1$ S(1) (ortho),
S(2) (para) and S(3) (ortho) lines are detected at high
signal-to-noise ratio and there is no indication for a deviation from
a ortho/para ratio of three in our sample. Together with
the above discussion, this argues against stellar UV-excitation.

\subsection{Shock Excitation}
A single shock model, where the various emission systems
are produced in one gas stream that cools after the shock,
does not predict the observed pressure differences
between the HII and H$_2$ regions.
A shock in cold gas, just strong enough to produce the 2000 K H$_2$
lines, would not produce the HII lines.  A stronger shock could produce
the HII lines, but, if the H$_2$ lines come from a cooler, post-shock
region, the pressure there would be lower than directly behind the
shock.  A third alternative, a shock ionizing gas previously warmed to
2000 K, would also produce higher pressure in the HII gas than in the
H$_2$ gas.

The consistent and low HII/H$_2$ line ratios 
severly restrict more complicated shock models.
As discussed in paper I, only a narrow range of shock
velocities, below $\sim 30$ km$^{-1}$, could produce the 
H$_2$ emission without excessive HII luminosity, while the
characteristic velocities in the centers of cD galaxies
are well above this.  Van der Werf, et al. (1993) discuss 
a similar situation in the merging galaxy pair NGC~6240
(see further the section on LINERS below).  They
propose, effectively, that dense molecular clouds collide
with a low density atomic or molecular ISM, 
forming a fast ionizing shock in the
low density medium and a slow, C-type shock in the molecular
medium.  They do not explicity calculate the ionized to
molecular line ratios in this case, and we find the
existence of such a low density, low temperature, medium
unlikely at the centre of the cooling flow.  

Alternatively, organized,
rotation-like, motions of the cloud ensemble may allow
the local cloud-cloud collisional velocities to be $\sim 30$ km $^{-1}$
while the large scale global motions are much higher.  In
this case the origin of the HII emission is unclear.
It cannot arise directly in the shocks (because of the
pressure argument) but might arise indirectly, for
example in a star forming region downstream.

While the HII and H$_2$ line systems could conceivably arise in
different regions from different mechanisms (e.g. HII emission
from a AGN narrow line region and H$_2$ lines from shocks in a
circumnuclear disk), the low variation of the line ratios from 
cluster to cluster argues against unrelated emission sources.

We conclude that our results can be explained by systems
of shocked clouds, but only under rather special
conditions, whose likelihood requires further investigation.
\subsection{Comparison to LINER spectra}
The relative strengths of H$_2$(1-0)S(1) and Br$\gamma$ in our spectra
are remarkably similar to those found in LINER galaxies (e.g., Larkin
et al., 1998), perhaps not surprisingly as the optical spectra of
cooling flows also have many similarities with LINERS (e.g.
VD97). Models for the excitation mechanisms in LINERS include
central AGN photoionization and ionization by an aging starburst
(effectively a mix of HII regions and supernova remnants, with a
relative deficiency in ionizing photons accounting for the faintness
of Br$\gamma$), and depend on the details of the near-IR spectrum
(Alonso-Herrero et al., 2000). What produces the H$_2$ vibrational
emission in LINERs is not clear, except in the class of LINERs where
shock-excitation dominates the observed spectrum, e.g., in NGC6240,
discussed above,
which has an optical spectrum which is typical for shock-excited
LINERs (e.g., Fosbury \& Wall 1979, Fried \& Schulz 1983, Heckman et
al.\ 1987). 
\section{Conclusions}
Our UKIRT observations detect multiple lines in six of the seven
cooling flux clusters observed.  In all cases except Abell 2029,
we detect at least four of the 1-0S series, sometimes some of the
2-1S series.  Paschen $\alpha$ is detected where the redshift
allows this, but Br$\gamma$ and $\delta$ are weak.

The ratio of molecular to ionized hydrogen line emission
is much higher than
those found in starburst regions or Seyfert galaxies, but
are similar to those in LINERS.  The ratios of
H$\alpha$ to Pa$\alpha$ do not indicate large absorption
of the former line by dust.

The 1-0S lines show near-LTE ratios, indicating densities well
above 10$^5$ cm$^{-3}$.  This implies pressures in the warm
gas that are 2 to 3 orders of magnitude higher than the
surrounding ionized gas, and rule out static equilibria.

The line ratio and pressure data argue against gas heating
by stellar UV emission.  X-ray heating in non-isobaric
conditions remains an option, as do shocks under 
highly constrained conditions.

\section*{Acknowledgments}
UKIRT is owned by the United Kingdom Particle Physics and Astronomy Research 
Council and is operated by the British-Dutch-Canadian Joint Astronomy
Centre at Hilo, Hawaii.

\bye